\documentclass[sigconf]{acmart}
\pdfoutput = 1
\renewcommand\footnotetextcopyrightpermission[1]{}
\usepackage{acronym}
\usepackage{amsmath}
\usepackage[english]{babel}
\usepackage{caption}
\usepackage[shortlabels,inline]{enumitem}
\usepackage{etoolbox}
\usepackage{graphicx}
\usepackage{hyperref}
\usepackage{bookmark}
\usepackage{ifthen}
\usepackage{cleveref}
\usepackage[utf8]{inputenc}
\usepackage{lipsum}
\usepackage{siunitx}
\usepackage{xcolor}
\usepackage[T1]{fontenc}

\usepackage{epstopdf}
\graphicspath{{./fig/}}
\usepackage[toc,symbols]{glossaries}
\usepackage{bm}
\usepackage{bbm}

\newtheorem{theo}{Theorem}

\newtheorem{prop}{Proposition}

\newcommand{\Figure}[1]{Fig.~\ref{#1}}
\newcommand{\Figures}[2]{Figs.~\ref{#1} and~\ref{#2}}

\newcommand{\Equation}[1]{\eqref{#1}}
\newcommand{\Equations}[2]{\eqref{#1} and~\eqref{#2}}
\newcommand{\Table}[1]{Table~\ref{#1}}

\newcommand{\Section}[1]{Section~\ref{#1}}

\newcommand{\Proposition}[1]{Proposition~\ref{#1}}

\newcommand\given[1][]{\:#1\vert\:}
\newcommand\prob[1]{\textnormal{Pr}\{{#1}\}}
\newcommand\fpdf[2][]{\textnormal{f}_{\mathrm{#1}}\left(#2\right)}
\newcommand{\poisDist}[2]{\ensuremath{\frac{{(#1)}^{#2} e^{-(#1)}}{#2!}}}
\newcommand{\pois}[1]{{\ensuremath{\textnormal{Pois}\left({#1}\right)}}}

\newcommand{\entropyBinary}[1]{\ensuremath{\mathrm{H}_{\textnormal{b}}\left({#1}\right)}}

\newcommand\todo[1]{\textcolor{red}{}}

\acrodef{3D}[3-D]{three-dimensional}
\acrodef{MC}{molecular communication}
\acrodef{OOK}{ON-OFF keying}
\acrodef{ISI}{inter-symbol interference}
\acrodef{IUI}{inter-user interference}
\acrodef{IR}{impulse response}
\acrodef{ML}{maximum likelihood}
\acrodef{wlog}[w.l.o.g.]{without loss of generality}
\acrodef{BER}{bit error rate}
\acrodef{SER}{symbol error rate}
\acrodef{BSC}{Binary Symmetric Channel}
\acrodef{CDF}{cumulated density function}
\acrodef{UCA}{uniform concentration assumption}
\acrodef{AWGN}{Additive White Gaussian Noise}
\acrodef{PBS}{particle-based simulation}
\acrodef{MLSE}{maximum likelihood sequence estimator}
\acrodef{VE}{viterbi equalizer}
\acrodef{MLE}{maximum likelihood estimator}
\acrodef{CIR}{channel impulse response}
\acrodef{CIRs}{channel impulse responses}
\acrodef{SNR}{Signal-to-Noise Ratio}
\acrodef{SDMA}{Space Division Multiple Access}
\acrodef{MCDMA}{Molecular Code Division Multiple Access}
\acrodef{ADMA}{Amplitude-Division Multiple Access}
\acrodef{MTDMA}{Molecular Time Division Multiple Access}
\acrodef{MDMA}{Molecular Division Multiple Access}
\acrodef{MIMO}{multiple-input multiple-output}
\acrodef{TDMA}{Time Division Multiple Access}
\acrodef{CDMA}{Code Division Multiple Access}
\acrodef{CSI}{channel state information}
\acrodef{TX}{transmitter}
\acrodef{TXs}{transmitters}
\acrodef{RX}{receiver}
\acrodef{RXs}{receivers}
\acrodef{ARE}{area rate efficiency}

\DeclareMathOperator\erf{erf}

\newcommand{\scaleSection}{\vspace*{-0.26cm}}
\newcommand{\scaleSubsection}{\vspace*{-0.25cm}}
\newcommand{\scaleSubsubsection}{\vspace*{-0.06cm}}

\newcommand{\scaleSubsectionBelow}{\vspace*{-0.1cm}}
\newcommand{\scaleSubsubsectionBelow}{\vspace*{-0.08cm}}
\newcommand{\scaleAlign}{\vspace*{-0.135cm}}

\newglossaryentry{ringNr}{%
  name = {\ensuremath{N_{\textnormal{rings}}}},
  description = {the current maximum ring number of interferers},
  type= symbols
}

\newglossaryentry{r0minHex}{%
  name = {\ensuremath{c}},
  description = {the distance between two adjacent transmitters in a hexagonal grid},
  type= symbols
}

\newglossaryentry{nrMol}{%
  name={\ensuremath{N_{\textnormal{m}}}},
  description={ the number of released molecules at the transmitter, which is equal for all the transmitters},
  type=symbols
}

\newglossaryentry{SRX}{%
  name={\ensuremath{S_{\mathrm{RX}}}},
  description={ the radius of a receiver},
  type=symbols
}
\newglossaryentry{SRX2}{%
  name={\ensuremath{S^2_{\mathrm{RX}}}},
  description={ squared radius of a receiver},
  type=symbols
}

\newglossaryentry{LRX}{%
  name={\ensuremath{L_{\mathrm{RX}}}},
  description={ the length of a receiver},
  type=symbols
}

\newglossaryentry{VRX}{%
  name={\ensuremath{V_{\mathrm{RX}}}},
  description={ the volume of a receiver, $\gls{VRX} = \gls{LRX} \pi \gls{SRX}^2$},
  type=symbols
}

\newglossaryentry{flow}{%
  name={\ensuremath{v}},
  description={ flow that points into the $z$ direction, $\gls{flow} = \gls{flow}_z$},
  type=symbols
}

\newglossaryentry{diffusion}{%
  name={\ensuremath{D}},
  description={ diffusion coefficient},
  type=symbols
}

\newglossaryentry{d}{%
  name={\ensuremath{d}},
  description={ the distance between transmitter and receiver plane},
  type=symbols
}

\newglossaryentry{nrIntU}{%
  name={\ensuremath{N_{\textnormal{IU}}}},
  description= the number of interfering users,
  type=symbols
}

\newglossaryentry{nrIntUG}{%
  name={\ensuremath{N_{\textnormal{IG}}}},
  description= the number of interfering groups,
  type=symbols
}

\newglossaryentry{nrTrans}{%
  name={\ensuremath{N_{\textnormal{TX}}}},
  description= the number of transmitters,
  type=symbols
}

\newglossaryentry{curTrans}{%
  name={\ensuremath{i}},
  description= {current transmitter with $\gls{curTrans} \in [1, \gls{nrTrans}]$},
  type=symbols
}

\newglossaryentry{nrRec}{%
  name={\ensuremath{N_\textnormal{RX}}},
  description= the number of transmitters,
  type=symbols
}

\newglossaryentry{curRec}{%
  name={\ensuremath{j}},
  description= {current receiver with $\gls{curRec} \in [1, \gls{nrRec}]$},
  type=symbols
}

\newglossaryentry{RX0}{%
  name={\ensuremath{\mathrm{RX}_0}},
  description= {current center receiver that will be the example receiver throughout the paper},
  type=symbols
}

\newglossaryentry{TX0}{%
  name={\ensuremath{\mathrm{TX}_0}},
  description= {current center transmitter that will be the example partner transmitter to \gls{RX0} throughout the paper},
  type=symbols
}

\newglossaryentry{samplingTime}{%
  name={\ensuremath{t_{\textnormal{s}}}},
  description= the time at which a receiver counts the currently receiving molecules,
  type=symbols
}

\newglossaryentry{CIR}{%
  name={\ensuremath{{\textnormal{CIR}_{\gls{curTrans},\gls{curRec}}(t)}}},
  description= number of molecules from transmitter \gls{curTrans} at receiver \gls{curRec} over time,
  type=symbols
}

\newglossaryentry{csmean}{%
  name={\ensuremath{\overline{c}_s}},
  description= {the expected number of molecules from the TX partner at sampling time $t = \gls{samplingTime}$ with $\gls{csmean} = \gls{nrMol} \gls{CIR}$ and $\gls{curTrans} = \gls{curRec}$},
  type=symbols
}

\newglossaryentry{cmeaniui}{%
  name={\ensuremath{\overline{\boldsymbol{c}}_{\textnormal{IUI}}}},
  description= {From the perspective of one receiver \gls{curRec} the expected number of molecules of all transmitters that are not the transmitter partner, $\gls{curTrans} \neq \gls{curRec}$ at sampling time $t = \gls{samplingTime}$ with $\gls{cmeaniui} = [\overline{c}_{1,\gls{curRec}}, \overline{c}_{2,\gls{curRec}}, \dots, \overline{c}_{\gls{curRec}-1,\gls{curRec}}, \overline{c}_{\gls{curRec}+1,\gls{curRec}}, \dots, \overline{c}_{\gls{nrTrans},\gls{curRec}}]$},
  type=symbols
}

\newglossaryentry{cmeaniuiTrans}{%
  name={\ensuremath{\overline{\boldsymbol{c}}^{\textnormal{T}}_{\textnormal{IUI}}}},
  description= {From the perspective of one receiver \gls{curRec} the expected number of molecules of all transmitters that are not the transmitter partner, $\gls{curTrans} \neq \gls{curRec}$ at sampling time $t = \gls{samplingTime}$ with $\gls{cmeaniui} = [\overline{c}_{1,\gls{curRec}}, \overline{c}_{2,\gls{curRec}}, \dots, \overline{c}_{\gls{curRec}-1,\gls{curRec}}, \overline{c}_{\gls{curRec}+1,\gls{curRec}}, \dots, \overline{c}_{\gls{nrTrans},\gls{curRec}}]$},
  type=symbols
}

\newglossaryentry{cmeann}{%
  name={\ensuremath{\overline{c}_n}},
  description= {the expected number of background noise molecules at sampling time $t = \gls{samplingTime}$},
  type=symbols
}

\newglossaryentry{cmeanij}{%
  name={\ensuremath{\overline{c}_{\gls{curTrans},\gls{curRec}}}},
  description= {the expected number of molecules from transmitter \gls{curTrans} at receiver \gls{curRec} at sampling time $t = \gls{samplingTime}$ with $\gls{cmeanij} = \gls{nrMol} \gls{CIR}$},
  type=symbols
}

\newglossaryentry{cmeaniuiScalar}{%
name={\ensuremath{\overline{c}_{\textnormal{IUI}}}},
description= {From the perspective of one receiver \gls{curRec} the expectation over the expected number of molecules of all transmitters that are not the transmitter partner, $\gls{cmeaniuiScalar} = \underset{\gls{siui}}{\mathbb{E}}\{\gls{siui}\gls{cmeaniui}^{\textnormal{T}}\}$},
type=symbols
}

\newglossaryentry{ss}{%
  name={\ensuremath{s_0}},
  description={From the perspective of one receiver \gls{curRec}, the current symbol of the TX partner, $\gls{curTrans} = \gls{curRec}$, with $\gls{ss} = \{0,1\}$, so there are two states for \gls{ss}},
  type=symbols
}
\newglossaryentry{ssHat}{%
  name={\ensuremath{\hat{s}_0}},
  description={From the perspective of one receiver \gls{curRec}, the current symbol of the TX partner, $\gls{curTrans} = \gls{curRec}$, with $\gls{ss} = \{0,1\}$, so there are two states for \gls{ss}},
  type=symbols
}

\newglossaryentry{Riui}{%
  name={\ensuremath{N_{\textnormal{IUI}}}},
  description={number of possible states of the interference coming from other users with  $\gls{Riui} = 2^{\gls{nrTrans}-1}$},
  type=symbols
}

\newglossaryentry{siui}{%
  name={\ensuremath{\boldsymbol{s}_{\textnormal{IUI}}}},
  description={From the perspective of one receiver \gls{curRec}, the current symbols of all transmitters that are not the transmitter partner, $\gls{curTrans} \neq \gls{curRec}$, with$\gls{siui} = [s_1, s_2, \dots, s_{\gls{curRec}-1}, s_{\gls{curRec}+1}, \dots, s_{\gls{nrTrans}}]$ and $s_i = \{0,1\}$, so there are $\gls{Riui} = 2^{\gls{nrTrans}-1}$ realisations possible},
  type=symbols
}

\newglossaryentry{cs}{%
  name={\ensuremath{c_s}},
  description= {a random number which represents the number of molecules from the TX partner at one receiver, $\gls{curTrans} = \gls{curRec}$ at sampling time \gls{samplingTime} with $\gls{cs} \sim \pois{\gls{ss}\gls{csmean} } = \fpdf{\gls{cs} \given \gls{ss} }$},
  type=symbols
}

\newglossaryentry{ciui}{%
  name={\ensuremath{c_{\textnormal{IUI}}}},
  description= {a random number which represents the number of molecules from the all transmitters that are not the transmitter partner, $\gls{curTrans} \neq \gls{curRec}$, at sampling time \gls{samplingTime} with $\gls{ciui} \sim \pois{\gls{siui}\gls{cmeaniui}^{\textnormal{T}} } = \fpdf{\gls{ciui} \given \gls{siui} }$. Note that \gls{siui} is a random binary vector},
  type=symbols
}

\newglossaryentry{cig}{%
  name={\ensuremath{\overline{c}_{\textnormal{IG}}}},
  description= {the expected number of molecules from transmitters of one group at receiver \gls{curRec} at sampling time $t = \gls{samplingTime}$},
  type=symbols
}

\newglossaryentry{cn}{%
  name={\ensuremath{c_{\textnormal{n}}}},
  description= {a random number which represents the number of noise molecules, so molecules from the background, at sampling time \gls{samplingTime} with $\gls{cn} \sim \pois{\gls{cmeann}} = \fpdf{\gls{cn}}$},
  type=symbols
}

\newglossaryentry{r}{%
  name={\ensuremath{r}},
  description= {the number of received molecules with $\gls{r} = \gls{cs} + \gls{ciui} + \gls{cn} \sim \pois{\gls{ss}\gls{csmean} + \gls{siui}\gls{cmeaniui}^{\textnormal{T}} + \gls{cmeann}} = \fpdf{r \given \gls{ss}, \gls{siui}}$ and therefore $\fpdf{\gls{r} \given \gls{ss}} = \sum_{\gls{siui}}\fpdf{\gls{r} \given \gls{ss}, \gls{siui}} \underbrace{\fpdf{\gls{siui}}}_{\frac{1}{\gls{Riui}}}$},
  type=symbols
}

\newglossaryentry{threshold}{%
  name={\ensuremath{\xi}},
  description= {a scalar value. If the received signal $\gls{r} \geq \gls{threshold}$ the detector will decide in favor of a $1$, i.e. $  \hat{\gls{ss}} = 1$},
  type=symbols
}

\newglossaryentry{groupl}{%
  name={\ensuremath{l}},
  description= {group index \gls{groupl} of the interfering group},
  type=symbols
}

\newglossaryentry{nrInIUIGroupl}{%
  name={\ensuremath{N_{\textnormal{IUI},\gls{groupl}}}},
  description= {the number of interfering transmitters in group \gls{groupl}},
  type=symbols
}

\newglossaryentry{nrInIUIGroup}{%
  name={\ensuremath{N_{\textnormal{IUI}}}},
  description= {the number of interfering transmitters in group \gls{groupl}},
  type=symbols
}

\newglossaryentry{errorProb}{%
  name={\ensuremath{P_e}},
  description= {the number of interfering transmitters in group \gls{groupl}},
  type=symbols
}

\newglossaryentry{effectiveRate}{%
  name = {\ensuremath{R_{\textnormal{eff}}}},
  description = {the total achievable rate of a spatial multiplexing system as described in the paper},
  type= symbols
}

\newglossaryentry{spatialRate}{%
  name = {\ensuremath{R_{\textnormal{loc}}}},
  description = {the rate due to the density of transmitters in a spatial multiplexing system, which is therefore a function of the distance between transmitters or receivers, i.e. the spatial packing},
  type= symbols
}

\setcopyright{acmcopyright}
\copyrightyear{}
\acmYear{}
\acmDOI{}

\acmConference[NANOCOM '21]{NANOCOM '21}{September 07--09, 2021}{Virtual Conference}

\begin{document}
\title{ Area Rate Efficiency in Molecular Communications}

\author{Lukas Brand\textsuperscript{*}, Sebastian Lotter\textsuperscript{*}, Vahid Jamali\textsuperscript{\textdagger}, and Robert Schober\textsuperscript{*}}
\affiliation{%
    \institution{\textsuperscript{*}Friedrich-Alexander University, Erlangen-Nuremberg, Germany
    \\
    \textsuperscript{\textdagger}Princeton University, Princeton, United States}
    \city{}
    \country{}
}
\email{lukas.brand@fau.de,sebastian.g.lotter@fau.de,jamali@princeton.edu,robert.schober@fau.de}

\begin{abstract}
\todo{Der todo makro ist nicht deaktiviert: Deakivieren, bevor das Dokument an jemanden gegben wird}We consider a multiuser diffusion-based \ac{MC} system where multiple spatially distributed \ac{TX}\acused{TXs}-\ac{RX}\acused{RXs}\todo{hier aufpassen, ob das noch gueltig ist} pairs establish point-to-point communication links employing the same type of signaling molecules. To realize the full potential of such a system, an in-depth understanding of the interplay between the spatial user density and \ac{IUI} and its impact on system performance in an asymptotic regime with large numbers of users is needed. In this paper, we adopt a \ac{3D} system model with multiple independent and spatially distributed point-to-point transmission links, where both the \ac{TXs} and \ac{RXs} are positioned according to a regular hexagonal grid, respectively. Based on this model, we first derive an expression for the \ac{CIRs}\acused{CIR} of all \ac{TX}-\ac{RX} links in the system. Then, we provide the \ac{ML} decision rule for the \ac{RXs} and show that it reduces to a threshold-based detector. We derive an analytical expression for the corresponding detection threshold which depends on the statistics of the \ac{MC} channel and the statistics of the \ac{IUI}. Furthermore, we derive an analytical expression for the \ac{BER} and the achievable rate of a single transmission link. Finally, we propose a new performance metric, which we refer to as \ac{ARE}, that captures the tradeoff between the user density and \ac{IUI}. The \ac{ARE} characterizes how efficiently given \ac{TX} and \ac{RX} areas are used for information transmission and is given in terms of bits per area unit. We show that there exists an optimal user density for maximization of the \ac{ARE}. Results from particle-based and Monte Carlo simulations validate the accuracy of the expressions derived for the \ac{CIR}, optimal detection threshold, \ac{BER}, and \ac{ARE}.

\end{abstract}

\maketitle
\pagestyle{plain}
\acresetall
\scaleSection
\section{Introduction}
In \ac{MC}, molecules are used to convey information. Possible industrial applications of \ac{MC} can be envisioned in bio-inspired robotics to improve search and rescue operations, in agriculture to understand and potentially control animal behavior, and in the oil and gas industry \cite{Farsad2016ACS}. In general, the data rate in \ac{MC} is low compared to conventional communication systems, which are based on electromagnetic waves. Furthermore, the particularities of \ac{MC} are often investigated for \ac{MC} systems which involve only a single transmission link, i.e., one \ac{TX}\acused{TXs} node that releases molecules and one \ac{RX}\acused{RXs}\todo{hier aufpassen, ob das wirklch die erste stelle ist} node that counts the received molecules. However, for practical applications, high data rates and large numbers of transmission links are desired. A common approach to increase the throughput of \ac{MC} systems is to decrease the symbol duration, which leads to \ac{ISI} \cite{Noel2014UENoiseISI, Tepekule2015ISIMItMC, Noel2014ISIMitByEnzymes}.

Besides the time dimension, the spatial dimension can also be exploited to increase the system throughput, which has received less attention. Existing works investigate the spatial dimension for example in the context of \ac{MIMO} \ac{MC} systems \cite{Meng2012MIMOMC, Damrath2018ArGain,Huang2019SpatMod,Gursoy2019IndMod,Rouzegar2019MIMO, Meng2012MIMOMC} or for large-scale \ac{MC} systems with many transmission nodes exploiting tools from \ac{3D} stochastic geometry \cite{Deng2017MU3D}.
An \ac{MC} \ac{MIMO} system comprises one \ac{TX} and one \ac{RX}, but the \ac{TX} and the \ac{RX} are connected to multiple spatially distributed release and reception sites, respectively. Hence, well known techniques such as spatial modulation for encoding \cite{Huang2019SpatMod,Gursoy2019IndMod}, and selection combining and zero forcing for decoding \cite{Rouzegar2019MIMO, Meng2012MIMOMC} are applicable. Performance benefits of \ac{MIMO} \ac{MC} systems, such as diversity gain and spatial multiplexing gain, are discussed in \cite{Meng2012MIMOMC, Damrath2018ArGain}.
In all existing studies on \ac{MIMO} \ac{MC} systems, the considered number of release/reception sites is quite low. To the best of the authors' knowledge, the largest such system was investigated in \cite{Gursoy2019IndMod}, which studied an $8 \times 8$ system.
Unlike the studies on \ac{MC} \ac{MIMO}, a large-scale \ac{MC} system with an asymptotically large number of transmitters is investigated in \cite{Deng2017MU3D}. There, a swarm of randomly placed transmitters simultaneously transmit \textit{the same bit sequence} to one receiver. Methods from stochastic geometry are utilized to derive the resulting received signal and the corresponding \ac{BER}.

In order to gain a fundamental understanding of the benefits of exploiting the spatial dimension in \ac{MC}, in this paper, we study an \ac{MC} system with a large number of \textit{independent} users, i.e., an asymptotic regime in terms of the number of \ac{TX} \textit{and} \ac{RX} nodes.
We raise the following research question: \textbf{For given transmit and receive areas, in which the \ac{TXs} and \ac{RXs} can be placed, how dense should the \ac{TXs} and \ac{RXs} be deployed to maximize the information transmission rate?}
Similar to other resources such as transmission time, the spatial resource allocated to different users should be optimized.
There exists a fundamental tradeoff between the achievable data rate of \textit{one} \ac{TX}-\ac{RX} link, which we refer to as user rate, and the link density. Increasing the density of \ac{TXs} and \ac{RXs} increases the number of independent transmission links per area unit. We refer to the number of links per area unit as spatial multiplexing rate with unit $\SI{}{[ 1 \per\metre\squared]}$. However, the molecules released from different \ac{TXs} may cause \ac{IUI} at the \ac{RX} units. Therefore, increasing the density of \ac{TXs} and \ac{RXs} increases the \ac{BER} and consequently decreases the user rate.

In order to analyze the aforementioned tradeoff, we propose a new performance metric for \ac{MC} systems which we refer to as \ac{ARE}. The \ac{ARE} is the product of the spatial multiplexing rate and the user rate and characterizes how efficiently a given area is used for information transmission and is given in terms of bits per area unit. To systematically address the question posed above, the following communication system is considered and evaluated.

We study a \ac{3D} multipoint-to-multipoint communication system comprising multiple independent, spatially distributed point-to-point transmission links and analyze the system in the asymptotic regime with a large number of users.  As in conventional wireless communications \cite{Andrews2011CellNetwork,Nasri2016HexagonalNetwork}, we define a location model for the positions of the \ac{TXs} and \ac{RXs} in the considered multi-link \ac{MC} setup, i.e., a fixed and predefined cellular structure. Furthermore, we assume that every \ac{TX} has exactly one \ac{RX} as desired communication partner. We evaluate the performance of the considered system by deriving an analytical expression for the \ac{ARE}. The main contributions of this work include:
\begin{enumerate}
  \item An analytical expression for the \ac{CIRs}\acused{CIR} of all \ac{TX}-\ac{RX} links for multipoint-to-multipoint molecule transmission via diffusion and directed flow is derived assuming point \ac{TXs} and transparent \ac{RXs} with cylindrical volume.
  \item Based on the \ac{CIR}, mathematical expressions for the optimal detection threshold of a threshold detector, the \ac{BER}, the user rate, the spatial multiplexing rate, and finally the \ac{ARE} are derived in the asymptotic regime for large numbers of users.
  \item By analyzing the \ac{ARE}, we demonstrate a fundamental tradeoff between the spatial multiplexing rate and the \ac{IUI} dependent user rate on the overall performance of the considered multipoint-to-multipoint transmission system. Our analysis shows that there exists an optimal number of transmission links per space unit which maximizes the \ac{ARE}.
\end{enumerate}

The remainder of this work is organized as follows. \Section{sec:model} introduces the system model. In \Section{sec:math}, the \ac{CIR} and the statistics of the \ac{IUI} are derived. The considered threshold detector and the performance metrics, namely the \ac{BER} and the \ac{ARE}, are analyzed in \Section{sec:performance}, and quantitatively studied and compared to particle-based and Monte Carlo simulations in \Section{sec:evaluation}. Finally, \Section{sec:conclusion} concludes this work.

\scaleSection
\section{System Model}
\label{sec:model}
\vspace*{-0.5ex}
\begin{figure}[!tbp]
  \begin{minipage}[t]{0.47\textwidth}
    \resizebox{85mm}{!}{
    \fontsize{8pt}{11pt}\selectfont
    \def\svgwidth{3.333in}
    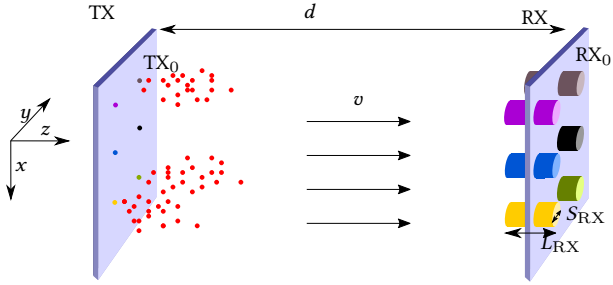}
    \caption{Unbounded system model with multiple, independent point-to-point transmission links (highlighted by different colors). The point \ac{TXs} are aligned within a predefined grid on a virtual \ac{TX} plane, shown in blue on the left, and release molecules (red dots) using \ac{OOK}. The molecules propagate by Brownian motion and directed flow and are detected by transparent volume \ac{RXs} which are aligned on a virtual \ac{RX} plane, depicted in blue on the right, on the same grid as the \ac{TXs}.}
    \label{fig:system_model}
  \end{minipage}
  \vspace{-0.5cm}
\end{figure}
\begin{figure}
     \begin{minipage}[t]{.47\textwidth}
       \hspace*{1.2cm}
       \resizebox{55mm}{!}{
       \rotatebox{90}{
       \fontsize{10pt}{13pt}\selectfont
       \def\svgwidth{3.333in}
       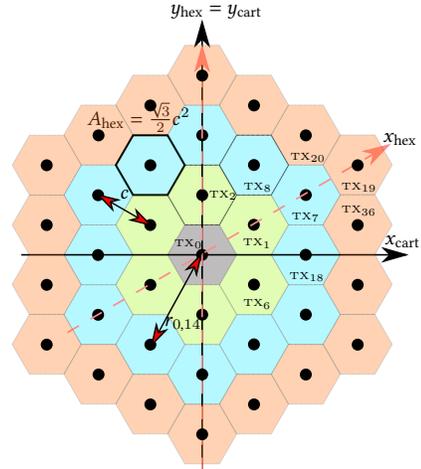}}
       \vspace*{-0.5cm}
       \caption{The point release positions are denoted as black dots within the virtual hexagonal grid with axes $x_{\textnormal{hex}}$ and $y_{\textnormal{hex}}$; the axes of the Cartesian coordinate system $x_{\textnormal{cart}}$ and $y_{\textnormal{cart}}$ are included as a reference. The distance $\gls{r0minHex}$ between two adjacent \ac{TXs} as well as the exemplary distance $r_{0,14}$ between the reference \ac{TX} in the center and $\ac{TX}_{14}$ are depicted by arrows. Multiple \ac{TXs} having a similar distance to \gls{RX0} align to rings of interferers which is highlighted with the same grid fill color, e.g., $\ac{TX}_1-\ac{TX}_6$ are in the green cells,  $\ac{TX}_7-\ac{TX}_{18}$ in the blue cells, and $\ac{TX}_{19}-\ac{TX}_{36}$ in the orange cells. }
       \label{graphic:System_model:Grid_hex}
     \end{minipage}
     \vspace{-0.5cm}
\end{figure}

We consider a \ac{3D} \ac{MC} environment with an infinite number of point \ac{TXs} and an infinite number of transparent \ac{RXs}, see \Figure{fig:system_model}. The \ac{TXs} are confined to a plane at $z=z_0$ and the \ac{RXs} are placed in a plane at $z=z_{\textnormal{R}}$. Both planes are virtual, i.e., are insubstantial, have infinite width and height, i.e., are infinite in $x$ and $y$ direction, and are separated by distance $\gls{d}$. In the following, these planes are referred to as TX-plane and RX-plane, respectively. Each \ac{TX} in the TX-plane is paired with the closest \ac{RX} in the RX-plane, i.e., each \ac{TX} wants to communicate with only one \ac{RX}. We assume that TX-plane and RX-plane are subdivided into an infinite number of hexagons, resulting in a hexagonal grid. The hexagonal grid is widely used for analyzing cellular architectures, e.g., in wireless communication networks \cite{Andrews2011CellNetwork,Nasri2016HexagonalNetwork}. Furthermore, hexagons are the closest shape to a circle that can still form a continuous grid, and therefore, guarantee the densest cellular packing. Each point \ac{TX} and the center of each \ac{RX} are in the center of a hexagon. The \ac{TX} and \ac{RX} hexagonal grids consist of regular, equally sized hexagons. The center points of the hexagons can be given in terms of an offset coordinate system $x', y'$ and are identical for the \ac{TX} plane and the \ac{RX} plane. The Euclidean distance between the center points of two adjacent cells is given by $\gls{r0minHex}$. For general center points, the distance to $(x',y') = (0,0)$ is $   \gls{r0minHex} \, \sqrt{ x'^2  + y'^2  + x' \; y' } $. The offset coordinate system $x', y'$ is related to the Cartesian coordinate system $x,y$ by $x = \gls{r0minHex} \, \frac{\sqrt{3}}{2} x', y = \gls{r0minHex} \, (y' + \frac{1}{2} x')$. The area of one hexagon is given by $A_\mathrm{hex} = \frac{\sqrt{3}}{2} \gls{r0minHex}^2$. The hexagonal grid structure is depicted in \Figure{graphic:System_model:Grid_hex}.
The considered arrangement ensures that for each \ac{TX}, the paired \ac{RX} has the shortest distance to it and that this distance is equal for all TX-RX pairs. We index the \ac{TXs} and \ac{RXs} as $\mathrm{TX}_i$ and $\mathrm{RX}_j$ with $i \in [0, \gls{nrTrans}-1]$ and $j \in [0, N_{\textnormal{RX}}-1]$, where $ \gls{nrTrans} \rightarrow \infty$ and $N_{\textnormal{RX}} \rightarrow \infty$, respectively. We note that the number of \ac{TXs} is equal to the number of \ac{RXs}, i.e., $\gls{nrTrans} = N_{\textnormal{RX}}$. Without loss of generality (w.l.o.g.)\acused{wlog} in this and the following sections, the analysis is done exemplary for the point-to-point transmission link between \gls{TX0} and \gls{RX0}, which are at the center of the \mbox{$x$-$y$ plane}, i.e., the center points are $(x_{\gls{TX0}},y_{\gls{TX0}}) = (x_{\gls{RX0}},y_{\gls{RX0}}) = (0,0)$, see \Figure{graphic:System_model:Grid_hex}. All other \ac{TXs}, i.e., $\mathrm{TX}_1, \mathrm{TX}_2, \dots, \mathrm{TX}_{\gls{nrTrans}-1}$, act as interferers to $\gls{TX0}$. All results are valid as well for the other TX-RX pairs due to the symmetry of the considered grids and the infinite extent of the \ac{TX}- and \ac{RX}-planes. We note that practical \ac{MC} systems will have finite dimensions, of course. However, considering a system with infinite dimensions is needed to gain fundamental insights for system design. Furthermore, considering infinite dimensions provides an upper bound on the \ac{IUI} for system models with a finite number of \ac{TXs} and \ac{RXs} as it constitutes the worst case scenario in terms of \ac{IUI}.

At the beginning of each symbol interval, each point $\mathrm{TX}_i$ releases a fixed non-zero number of molecules, \gls{nrMol}, or zero molecules, representing binary symbols $s_i = 1$ and $s_i = 0$, respectively, i.e., the \ac{TXs} use \acf{OOK} modulation \cite{Jamali2019ChannelMF}. We assume that the binary values $0$ and $1$ are equiprobable, i.e., $\prob{s_i = 1} = 0.5$\\ $= \prob{s_i = 0}$. All molecules considered in this setup are of the same type. The symbol intervals are synchronized between the \ac{TXs}, i.e., the \ac{TXs} transmit at the same time instants. We assume that the release, propagation, and reception of all molecules are independent from each other. The movement of the released molecules from the \ac{TXs} to the \ac{RXs} is affected by diffusion and a uniform flow in $z$ direction with velocity $\gls{flow} = \gls{flow}_z$. We neglect additional effects due to external forces, degradation, and interactions between molecules. Each \ac{RX} counts the number of molecules within its volume at a fixed sampling time \gls{samplingTime}, i.e., we consider transparent \ac{RXs} in this work. The sampling time \gls{samplingTime} at \gls{RX0} is chosen such that it coincides with the time instant where the \ac{CIR} between \gls{TX0} and \gls{RX0} has its peak. All \ac{RXs} have the same size and a cylindrical shape, and can be characterized in terms of their radius $\gls{SRX}$ and length $\gls{LRX}$. We chose cylindrical receivers as this enables an independent scaling of the receiver size in the \mbox{$xy$}-plane via \gls{SRX} and the $z$-axis via \gls{LRX}. The observed number of molecules, $\gls{r}$, within \gls{RX0} comprises the molecules originating from \gls{TX0}, $\gls{cs}$, and molecules received from other \ac{TXs}, $\gls{ciui}$, which constitute \ac{IUI}. Hence, the received number of molecules at the sampling time at \gls{RX0} is given by
\scaleAlign
\begin{align}
  \gls{r} = \gls{cs} + \gls{ciui} \;.
  \label{eq:system_model_received_molecules}
\end{align}
We note that multiple consecutive emissions of molecules may cause ISI if the symbol duration is too short. However, since the focus of this paper is the spatial dimension, we assume that ISI is negligible which is a valid assumption if the symbol duration is sufficiently large.

\scaleSection
\section{Analytical Channel Model}
\label{sec:math}
In this section, we first derive an expression for the \ac{CIRs} of all transmission links that involve receiver \gls{RX0}. We note that the derived \ac{CIR} expression is also valid for all other \ac{TX}-\ac{RX} pairs due to the considered symmetric system model. Next, we analyze the distributions of the information and the interfering molecules.
\scaleSubsection
\subsection{Channel Impulse Response}\label{math_sec:cir}
\scaleSubsectionBelow
We consider an unbounded environment with constant advection along the $z$-axis with velocity $\gls{flow} = \gls{flow}_z$ and diffusion coefficient $\gls{diffusion}$.
We derive the expected number of molecules observed at $\gls{RX0}$ normalized to the instantaneous release of one molecule by $\ac{TX}_{\gls{curTrans}}$ at position  $\boldsymbol{p}_{0,\gls{curTrans}} = (x_{0,\gls{curTrans}}, y_{0,\gls{curTrans}}, z_{0,\gls{curTrans}})$ and time $t_{0,\gls{curTrans}} = 0$ as a function of time, which we refer to as \ac{CIR}.
In the following proposition, we provide an analytical expression for the \ac{CIR} in terms of an infinite sum.
\vspace*{-0.2cm}
\begin{prop}\label{prop:cir}
The expected number of particles at the cylindrical and transparent receiver \gls{RX0} with dimensions $0 \leq r= \sqrt{ x^2 + y^2} \leq \gls{SRX}$, $0 \leq \varphi < 2 \pi$, and $z_{\mathrm{S}} < z < z_{\mathrm{E}}$, due to the release of one molecule at transmitter $\ac{TX}_{\gls{curTrans}}$ is given by
\scaleAlign
  \begin{align}
  \textnormal{CIR}_{\gls{curTrans},0}(t) &= \frac{1}{2} \left(\erf\left(\frac{z_{0} + \gls{flow} t - z_{\mathrm{S}}}{\sqrt{4 D t}}\right)- \erf\left(\frac{z_{0} +\gls{flow} t - z_{\mathrm{E}}}{\sqrt{4 \gls{diffusion} t}}\right)\right) \nonumber \\
  & \qquad \times e^{-\frac{r_{0,\gls{curTrans}}^2}{4 \gls{diffusion} t}} \sum_{k=0}^{k_{\mathrm{max}} = \infty} \frac{(\frac{r_{0,\gls{curTrans}}^2}{4 \gls{diffusion} t})^k}{(k!)^2} \gamma(k+1,\frac{\gls{SRX2}}{4 \gls{diffusion} t}) \;,
  \label{eq:math_ir_general}
  \end{align}
  where $r_{0,\gls{curTrans}}^2 = x_{0,\gls{curTrans}}^2 + y_{0,\gls{curTrans}}^2$. Here, $\varphi$ denotes the angle in the $xy$-plane in relation to the positive $x$-axis. Furthermore, $\erf(x)$, $\mathrm{I}_0(x)$, and $\gamma(a,x)$ denote the Gaussian error function, the zeroth order modified Bessel function of the first kind, and the lower incomplete Gamma function, respectively.
\end{prop}
\vspace*{-0.3cm}
\begin{proof}
The molecule concentration $C(x, x_{0,\gls{curTrans}}, y, y_{0,\gls{curTrans}}, z, z_{0,\gls{curTrans}}, t, t_{0,\gls{curTrans}})$ for the case considered in \Proposition{prop:cir} is given in \cite[eq. (18)]{Jamali2019ChannelMF}. \\$C(x, x_{0,\gls{curTrans}}, y, y_{0,\gls{curTrans}}, z, z_{0,\gls{curTrans}}, t, t_{0,\gls{curTrans}})$ can be transformed to a cylindrical coordinate system by \mbox{$x = r \cos(\varphi)$}, \mbox{$y = r \sin(\varphi)$}, \mbox{$x_{0,\gls{curTrans}} = r_{0,\gls{curTrans}} \cos(\varphi_{0,\gls{curTrans}})$}, \mbox{$y_{0,\gls{curTrans}} = r_{0,\gls{curTrans}} \sin(\varphi_{0,\gls{curTrans}})$}, \mbox{$z = z$}. Assuming $t_{0,\gls{curTrans}} = 0$ and noting that $z_{0,\gls{curTrans}} = z_0$ is valid for all \ac{TXs}, we obtain:
  \scaleAlign
  \begin{align}
    &C_{\mathrm{c}}(r, r_{0,\gls{curTrans}}, \varphi, \varphi_{0,\gls{curTrans}}, z, z_{0}, t) = \frac{1}{(4 \pi \gls{diffusion} t)^{3/2}}\nonumber \\
    & \times \exp\left(-\frac{r^2+r_{0,\gls{curTrans}}^2 - 2 r_{0,\gls{curTrans}} r \cos(\varphi-\varphi_{0,\gls{curTrans}}) + (z-z_{0}-\gls{flow} t)^2}{4 \gls{diffusion} t} \right)  \;.
  \end{align}
  Furthermore, since the center of the circular cross section of the receiver is located at $r=0$, the expected number of counted particles is obtained as
  \scaleAlign
  \begin{align}
    &\textnormal{CIR}_{\gls{curTrans},0}(t)= \int \limits_0^{\gls{SRX}} \int \limits_0^{2 \pi} \int \limits_{z =z_{\mathrm{S}}}^{z_{\mathrm{E}}}  C_{\mathrm{c}}(r, r_{0,\gls{curTrans}}, \varphi, \varphi_{0,\gls{curTrans}}, z, z_{0}, t) r \, \mathrm{d} z  \,\mathrm{d}  \varphi   \,\mathrm{d}  r \nonumber \\
    &\overset{(a)}{=} \frac{1}{4 \gls{diffusion} t} \left(\erf\left(\frac{z_{0} + \gls{flow} t - z_{\mathrm{S}}}{\sqrt{4 \gls{diffusion} t}}\right)- \erf\left(\frac{z_{0} +\gls{flow} t - z_{\mathrm{E}}}{\sqrt{4 \gls{diffusion} t}}\right)\right) \nonumber \\
    &\qquad \times \int \limits_0^{\gls{SRX}}  \mathrm{I}_0\left(\frac{r_{0,\gls{curTrans}} r}{2 \gls{diffusion} t}\right) \exp\left(-\frac{r^2+r_{0,\gls{curTrans}}^2}{4 \gls{diffusion} t} \right) r \,\mathrm{d}  r \nonumber  \\
    &\overset{(b)}{=} \frac{1}{4 \gls{diffusion} t} \left(\erf\left(\frac{z_{0} + \gls{flow} t - z_{\mathrm{S}}}{\sqrt{4 \gls{diffusion} t}}\right)- \erf\left(\frac{z_{0} +\gls{flow} t - z_{\mathrm{E}}}{\sqrt{4 \gls{diffusion} t}}\right)\right) \nonumber \\
    & \times \exp\left(-\frac{r_{0,\gls{curTrans}}^2}{4 \gls{diffusion} t} \right) \mkern-5mu \int \limits_0^{\gls{SRX}} \, \sum_{k=0}^{k_{\mathrm{max}} = \infty}\mkern-15mu \exp\left(-\frac{r^2}{4 \gls{diffusion} t} \right) \frac{\left(\frac{r_{0,\gls{curTrans}} r}{4 \gls{diffusion} t}\right)^{2 k} }{(k!)^2} r \,\mathrm{d}  r \;,
    \label{eq:math_section:CIR}
  \end{align}
  where we exploited $(a)$ $\erf(x) = \frac{2}{\sqrt{\pi}} \int_0^x \exp\left(-y^2 \right)  \,\mathrm{d}  y$ and $  \mathrm{I}_0(x) = \frac{1}{\pi} \int_0^{\pi} \exp\left(x \cos(\varphi)\right) \,\mathrm{d}  \varphi,$ and $(b)$ the series expansion of $ \mathrm{I}_0(x)$ \cite[eq. (9.6.10)]{abramowitz1964handbook}, and $\gamma(a,x) = \int_0^x y^{a-1} \exp(-y) \,\mathrm{d}  y$ to obtain \Equation{eq:math_ir_general}.
\end{proof}
Eq. \Equation{eq:math_ir_general} shows that the expected number of particles received from different \ac{TX}s depends on their release positions, characterized by $r_{0,\gls{curTrans}}$. All other parameters in \Equation{eq:math_ir_general}, e.g., the $z$-position of the \ac{TXs}, the diffusion coefficient $\gls{diffusion}$ and the flow velocity \gls{flow} are equal for all \ac{TXs}.
We note that, as will be shown in \Section{sec:CIR_verification}, the infinite sum over $k$ in \Equation{eq:math_ir_general} can be truncated to a small number of terms (e.g., $20$) without compromising its accuracy.
\scaleSubsection
\subsection{Statistical Model}\label{ssSec:IM}
\scaleSubsectionBelow
We now describe the underlying models and associated distributions of $\gls{cs}$ and $\gls{ciui}$.
\scaleSubsubsection
\subsubsection{Information Molecules}\label{ssSec:IM}
\scaleSubsubsectionBelow
Information carrying molecules intended for \gls{RX0} are released for signaling $\gls{ss}=1$ at \gls{TX0}. At time instant $t_m$, a corresponding particle is received with probability $\textnormal{CIR}_{0,0}(t=t_m)$ at \gls{RX0}. The expected (non-normalized) number of observed information molecules at sampling time $\gls{samplingTime}$ is given by $ \gls{csmean} =  \gls{nrMol} \; \textnormal{CIR}_{0,0}(\gls{samplingTime}).$
We statistically model the reception of the released particles by a Poisson distribution, i.e.,
\scaleAlign
\begin{align}
  \gls{cs} \sim \pois{\gls{ss}\gls{csmean} } = \fpdf[s]{\gls{cs} \given \gls{ss} } \;,
  \label{equation:math_sec:ss_pois}
\end{align}
which is a valid approximation for typical \ac{MC} scenarios \cite{Jamali2019ChannelMF}. Here, $\fpdf[s]{\gls{cs} \given \gls{ss} }$ denotes the distribution of $\gls{cs}$ conditioned on \gls{ss}.
\scaleSubsubsection
\subsubsection{Interference Molecules from other Transmitters}\label{ssSec:IUIMol}
\scaleSubsubsectionBelow
Besides the information molecules, molecules emitted by $\mathrm{TX}_1, \mathrm{TX}_2, \mathrm{TX}_3, \dots, \\ \mathrm{TX}_{\gls{nrTrans}-1}$, i.e., \ac{TX}s belonging to other TX-RX pairs, are received as \ac{IUI} at $\gls{RX0}$. These molecules are not distinguishable from the information molecules. Thereby, \ac{IUI} degrades the detection of the signal from \gls{TX0} at \gls{RX0} \cite{Noel2014UENoiseISI}. We note that $\ac{TX}_{\gls{curTrans}}$, $\gls{curTrans} \neq 0$, causes interference at \gls{RX0} only for $s_{\gls{curTrans}} = 1$ as no molecules are released for $s_{\gls{curTrans}} = 0$. Hence, which of the interferers are active, i.e., send a $1$, and the Brownian motion of the molecules both introduce randomness. So, for a complete analysis, all possible combinations of active \ac{TX}s as well as the randomness and statistics of the molecule movement have to be taken into account.

From the perspective of \gls{RX0}, the expected \ac{IUI} at sampling time $t = \gls{samplingTime}$ can be collected in vector $\gls{cmeaniui} = [\overline{c}_{1}, \overline{c}_{2}, \dots, \overline{c}_{\gls{nrTrans}-1}] $ with  $ \overline{c}_{i} =  \gls{nrMol} \; \textnormal{CIR}_{i,0}(t=\gls{samplingTime})$.
The received number of molecules from each interferer can be characterized by a Poisson distribution \cite{Jamali2019ChannelMF}. The molecule releases at the different \ac{TX}s are independent of each other. As \gls{RX0} counts all molecules independent of their origin, the sum over all Poisson distributed interfering molecules follows again a Poisson distribution, i.e.,
\scaleAlign
\begin{align}
   \gls{ciui} \sim \pois{\gls{siui}\gls{cmeaniuiTrans} } = \fpdf[IUI]{\gls{ciui} \given \gls{siui} } \;,
   \label{equation:math_sec:siui_pois}
\end{align}
where $\fpdf[IUI]{\gls{ciui} \given \gls{siui} }$ denotes the distribution of \gls{ciui} conditioned on vector $\gls{siui} = [s_1, s_2, \dots, s_{\gls{nrTrans}-1}]$, which contains the symbols emitted by the interfering \ac{TXs}. There are $\gls{Riui} = 2^{\gls{nrTrans}-1}$ possible realizations of \gls{siui}, i.e., different \ac{IUI} states, which are equiprobable due to the equiprobable binary transmission symbols. Note that $\gls{Riui}$ grows exponentially in $\gls{nrTrans}$ and we assumed $\gls{nrTrans} \rightarrow \infty$.
However, we will later show numerically that the \ac{IUI} can be accurately approximated by taking into account only a finite number of interfering \ac{TXs}. The \ac{IUI} is dominated by the strongest interferers, which correspond to the \ac{TXs} nearest to \gls{TX0}. From \Figure{graphic:System_model:Grid_hex}, it can be seen that the indexes of the \ac{TXs} are chosen such that, for increasing index number, the distance between \gls{TX0} and $\mathrm{TX}_{\gls{curTrans}}$ is monotonically non-decreasing in \gls{curTrans}. Therefore, the \textit{strongest} interferers are always included if \gls{nrTrans} is truncated to a finite number. We note that the actual number of \ac{TXs} necessary to accurately approximate the system behavior depends on the channel parameters, see \Section{eval:BER_ARE_Results}.

\scaleSection
\section{Symbol Detection and Performance Analysis}
\label{sec:performance}
In this section, we first derive the optimal \ac{ML} decision rule and the \ac{BER} of one TX-RX pair. Then, we provide the user rate defined as the achievable information transmission rate of an individual transmission link, and the spatial multiplexing rate which depends on the spatial density of the \ac{TX}-\ac{RX} pairs. Finally, we introduce the \ac{ARE} to characterize the performance of the entire system.
\scaleSubsection
\subsection{Optimal ML Detector}\label{math_sec:ssMLDet}
\scaleSubsectionBelow
For detection, we assume all \ac{CIR}s in the system are known but the activity of the interfering \ac{TXs} is only statistically known. Therefore, the \ac{ML} estimate, \gls{ssHat}, is given by
\scaleAlign
\begin{align}
\gls{ssHat} &= \underset{\gls{ss} \in \{0,1\}}{\text{argmax}} \; \fpdf[r]{\gls{r} \given \gls{ss}}  = \underset{\gls{ss} \in \{0,1\}}{\text{argmax}} \;  \underset{\gls{siui}}{\mathbb{E}} \{\fpdf[r]{\gls{r} \given \gls{ss}, \gls{siui}}\} \label{eq:math_section_decision_metric}\;\\
  &= \begin{cases}
      1 , \; \text{if} \; \frac{\sum\limits_{\gls{siui} \in M }  \poisDist{\gls{csmean} + \gls{siui}\gls{cmeaniuiTrans}}{\gls{r}}}{\sum\limits_{\gls{siui} \in M}  \poisDist{\gls{siui}\gls{cmeaniuiTrans}}{\gls{r}}} \geq 1 \\
      0 , \;\text{otherwise}
  \end{cases}\;,
  \label{eq:math_section_ML}
\end{align}
where $\fpdf[r]{\gls{r} \given \gls{ss}}$ denotes the distribution of $\gls{r}$ conditioned on \gls{ss}, which is a Poisson distribution with mean $\gls{ss}\gls{csmean}  + \gls{siui}\gls{cmeaniuiTrans}$ as both \gls{cs} and \gls{ciui} are Poisson distributed, cf. \Equation{eq:system_model_received_molecules}. Here, $\underset{\gls{siui}}{\mathbb{E}}\{\cdot\}$ and $M = \{0,1\}^{\gls{nrTrans}-1}$ denote the expectation over the \ac{IUI} and the set of interference symbols, respectively. The \ac{ML} detection in \Equation{eq:math_section_ML} is computationally complex as all possible realizations of the \ac{IUI} have to be taken into account and has to be computed for every received \gls{r}. Note that evaluating \Equation{eq:math_section_ML} is only feasible for a finite number of interferers, see \Section{ssSec:IUIMol}.
In the following, we show that \Equation{eq:math_section_ML} can be equivalently realized by a threshold detector employing a pre-calculated threshold. Threshold detection is preferable as the threshold for a given setup has to be computed only once offline and then can be used throughout the transmission. Therefore, the computational cost for online data detection is reduced.
\vspace*{-0.2cm}
\begin{theo}
  The \ac{ML} detection problem given in \Equation{eq:math_section_ML} can be written equivalently as a threshold detection
\scaleAlign
  \begin{align}
    \gls{ssHat} &=
        \begin{cases}
          1 , \; \text{if} \; \gls{r} \geq  \gls{threshold}\\
          0 , \;\text{otherwise}
        \end{cases}\;
        \label{eq:math_section:threshold}
  \end{align}
  with threshold value $\gls{threshold}$ obtained by
\scaleAlign
  \begin{align}
    \gls{threshold} &= \text{min} \left\{ \Theta \in \mathbb{N} \;|  \sum_{\gls{siui} \in M}{(\gls{csmean} + \gls{siui}\gls{cmeaniuiTrans})}^{\Theta} e^{-(\gls{csmean} + \gls{siui}\gls{cmeaniuiTrans})} \right. \nonumber \\[-0.25cm]
    & \hspace{1.5cm} \left. \geq \sum_{\gls{siui} \in M} {(\gls{siui}\gls{cmeaniuiTrans})}^{\Theta} e^{-(\gls{siui}\gls{cmeaniuiTrans})}  \right\} \;.
    \label{eq:math_section_threshold_definition}
  \end{align}
\end{theo}
\vspace*{-0.3cm}
\begin{proof}
  Due to space limitation, we provide only a sketch of the proof. The existence of a unique threshold level \gls{threshold} can be proved by showing that the fraction in \Equation{eq:math_section_ML} is monotonically increasing in $\gls{r}$, i.e., for low values of \gls{r}, i.e., $\gls{r} <  \gls{threshold}$, the fraction is always smaller than $1$, and for large values of \gls{r}, i.e., $\gls{r} \geq  \gls{threshold}$, the fraction is always at least equal to $1$. This can be proved using the same steps as in a similar proof in \cite[Appendix]{Jamali2018NonCoherentDF}.
\end{proof}
\scaleSubsection
\subsection{Bit Error Rate of one \ac{TX}-\ac{RX} Pair}
\scaleSubsectionBelow
The \ac{BER} can be expressed as
\scaleAlign
\begin{align}
    \gls{errorProb}&= \underset{\gls{ss}}{\mathbb{E}}\{\underset{\gls{siui}}{\mathbb{E}}\{ \gls{errorProb}(\gls{ssHat} \given \gls{siui}, \gls{ss}) \}\} \nonumber \\
    &= \sum_{\gls{ss}} \sum_{\gls{siui}} \gls{errorProb}(\gls{ssHat} \given \gls{siui}, \gls{ss}) \fpdf{\gls{siui}} \fpdf{\gls{ss}} \;,
\end{align}
where $\gls{errorProb}(\gls{ssHat} \given \gls{siui}, \gls{ss})$ is the error probability conditioned on given transmitted symbol $\gls{ss}$ and interference vector $\gls{siui}$. In the following, we derive the \ac{BER} for the proposed threshold detector.
\vspace*{-0.2cm}
\begin{prop}
  For the proposed threshold detector, \gls{errorProb} can be expressed as
\vspace*{-0.25cm}
  \begin{align}
     \gls{errorProb} &= \frac{1}{2}  \underbrace{\frac{1}{\gls{Riui}}  \sum\limits_{\gls{siui} \in M} \mathcal{Q}(\gls{threshold},\gls{csmean} + \gls{siui}\gls{cmeaniuiTrans})}_{ q = \prob{\gls{ssHat} = 0 \given \gls{ss} = 1}} \nonumber \\
     &\quad + \frac{1}{2} \underbrace{\frac{1}{\gls{Riui}}  \sum\limits_{\gls{siui} \in M}   (1 -  \mathcal{Q}(\gls{threshold}, \gls{siui}\gls{cmeaniuiTrans}))}_{p = \prob{\gls{ssHat} = 1 \given \gls{ss} = 0}} \;,
    \label{eq:performance:error_prob}
  \end{align}
  where $\mathcal{Q}(a,b) $, $q$, and $p$ denote the regularized Gamma function, the error probability for $\gls{ss} = 1$, and the error probability for $\gls{ss} = 0$, respectively.
\end{prop}
\vspace*{-0.3cm}
\begin{proof}
  $\gls{errorProb}$ can be derived as follows:
\scaleAlign
  \begin{align}
    \gls{errorProb}&\overset{(a)}{=}  \frac{1}{\gls{Riui}}  \sum\limits_{\gls{siui} \in M} \left( \frac{1}{2} \prob{ \gls{r} < \gls{threshold} \given \gls{siui}, \gls{ss} = 1}\right.\nonumber \\[-0.25cm]
     &\hspace{2.5cm} \left. + \frac{1}{2} \prob{\gls{r} \geq \gls{threshold} \given \gls{siui}, \gls{ss} = 0}\right) \nonumber \\
    &\overset{(b)}{=} \frac{1}{\gls{Riui}}  \sum\limits_{\gls{siui} \in M} \left( \frac{1}{2}  \sum_{\gls{r} = 0}^{\gls{threshold}-1}  \poisDist{\gls{csmean} + \gls{siui}\gls{cmeaniuiTrans}}{\gls{r}}  \right. \nonumber \\
    & \hspace{2.5cm}\left.  + \frac{1}{2} \biggl(1 - \sum_{\gls{r} = 0}^{\gls{threshold}-1} \poisDist{\gls{siui}\gls{cmeaniuiTrans}} {\gls{r}}\biggr)\right) \;,
  \end{align}
  where we exploit in $(a)$ the threshold detection rule \Equation{eq:math_section:threshold}, in $(b)$ the fact that $\gls{r}$ is Poisson distributed, and finally the \ac{CDF} of a Poisson distribution $\sum_{k = 0}^{x-1} \poisDist{\lambda}{k} = \frac{\Gamma( x ,\lambda)}{\Gamma(x )} = \mathcal{Q}(x ,\lambda), \; \mathrm{with}\; x>0$ to obtain \Equation{eq:performance:error_prob}. Here, $\Gamma(a,b)$, $\Gamma(a)$, and $\frac{\Gamma(a,b)}{\Gamma(a)} = \mathcal{Q}(a,b)$ denote the upper incomplete Gamma function, the Gamma function, and the regularized Gamma function, respectively.
\end{proof}
\scaleSubsection
\subsection{Area Rate Efficiency of \ac{MC} Systems}
\scaleSubsectionBelow
Now, we provide the achievable information transmission rate of an individual transmission link, the spatial multiplexing rate, and finally the \ac{ARE} which involves both aforementioned rates.
\scaleSubsubsection
\subsubsection{User Rate}
\scaleSubsubsectionBelow
As the transmission is binary, i.e., $\gls{ss}, \gls{ssHat} \in \{0,1\}$, we model the point-to-point transmission link as a binary channel with achievable data rate \cite[eq. (9.7)]{Mackay2003information}
\scaleAlign
\begin{align}
  R_{\textnormal{SISO}} = \mathrm{I}(\gls{ss};\gls{ssHat}) = \mathrm{H}(\gls{ssHat}) - \mathrm{H}(\gls{ssHat} \given \gls{ss}) \;,
  \label{eq:performacne_user_rate}
\end{align}
which we refer to as user rate. Here, $\gls{ss}$, $\gls{ssHat}$, $\mathrm{I}(\cdot;\cdot)$, and $\mathrm{H}(\cdot)$ denote the channel input, the channel output, the mutual information, and the entropy function, respectively, where $\mathrm{H}(\gls{ssHat}) = -\frac{1}{2}((1-p+q) \log_2(\frac{1}{2} (1-p+q)) + (1-q+p) \log_2(\frac{1}{2} (1-q+p)))$ \cite[eq. (2.35)]{Mackay2003information} and $\mathrm{H}(\gls{ssHat} \given \gls{ss}) = -\frac{1}{2} ((1-p) \log_2(1-p) + p \log_2(p) + q \log_2(q) + (1-q) \log_2(1-q))$ \cite[eq. (8.4)]{Mackay2003information}.
Two extreme cases of the binary channel are the Z-Channel and the \ac{BSC} \cite{Mackay2003information}. A Z-Channel is characterized by error-free transmission for $\gls{ss} = 0$, i.e., $p = 0$, and possible errors for $\gls{ss} = 1$ and is an appropriate approximation, when the \ac{IUI} approaches zero.
On the other hand, for large \ac{IUI},  the difference between the two possible transmission errors $p$ and $q$ approaches zero. Hence, the \ac{BSC}, which assumes $p = q$, accurately approximates the binary channel for cases where \gls{r} in \Equation{eq:system_model_received_molecules} is dominated by \ac{IUI}. We note that in a \ac{BSC} the mutual information \cite{Mackay2003information} is maximized by equiprobable binary input and \Equation{eq:performacne_user_rate} simplifies to the channel capacity of a \ac{BSC}, i.e., $R_{\textnormal{SISO}} = C_{\textnormal{\ac{BSC}}} = [1-\entropyBinary{\gls{errorProb}}]$ \cite{Mackay2003information}. Here, $\entropyBinary{\gls{errorProb}} $ denotes the binary entropy function for a given error probability $\gls{errorProb} = p = q$, c.f. \Equation{eq:performance:error_prob}. However, to ensure the accuracy for all levels of \ac{IUI}, we employ \Equation{eq:performacne_user_rate} for our analysis in \Section{sec:evaluation}.
\scaleSubsubsection
\subsubsection{Spatial Multiplexing Rate}
\scaleSubsubsectionBelow
We define the spatial multiplexing rate as the number of transmission links per unit of space, i.e., $ \gls{spatialRate} = \frac{1}{A_{\textnormal{cell}}}$ with unit $\SI{}{[ 1 \per\metre\squared]}$, where $A_{\textnormal{cell}}$ is the area reserved for one TX-RX pair, i.e., one hexagon. Note that the smaller $A_{\textnormal{cell}}$, the higher the spatial multiplexing rate.
\scaleSubsubsection
\subsubsection{Area Rate Efficiency}
\scaleSubsubsectionBelow
We define the \ac{ARE} based on $R_{\textnormal{SISO}}$ and $\gls{spatialRate}$ as follows
\vspace*{-0.2cm}
\begin{align}
  \gls{effectiveRate} &= R_{\textnormal{SISO}} \gls{spatialRate} =   \frac{1}{A_{\textnormal{cell}}} [\mathrm{H}(\gls{ssHat}) - \mathrm{H}(\gls{ssHat} \given \gls{ss}) ] \,
\label{eq:performance:effRate}
\end{align}
with unit $\SI{}{[ \textnormal{bit} \per\metre\squared]}$. Note that the two rates, $R_{\textnormal{SISO}}$ and $\gls{spatialRate}$, have different dependencies on the density of the \ac{TXs} and \ac{RXs} in the system. In particular, increasing the density of the TX-RX pairs increases $\gls{spatialRate}$, but decreases $R_{\textnormal{SISO}}$ as the \ac{BER} increases, and vice versa. Hence, the \ac{ARE} comprises the single link performance in terms of $R_{\textnormal{SISO}}$ \textit{and} the area usage efficiency in terms of $\gls{spatialRate}$, and therefore provides a useful performance metric to gain insight into how efficiently given \ac{TX} and \ac{RX} areas are exploited for maximization of the overall information transmission of a system.

\scaleSection
\section{Performance Evaluation}
\label{sec:evaluation}
In this section, we first explain the simulation setup. Then, we evaluate the analytical expression for the \ac{CIR} in \Equation{eq:math_ir_general} and compare it to results from \ac{PBS}. Finally, the analytical expressions for the \ac{BER} and the \ac{ARE} are evaluated.
%
%
\scaleSubsection
\subsection{Choice of Parameters, Particle-Based Simulation, and Monte Carlo Simulation}\label{section:evaluation:PBSAndParam}
\scaleSubsectionBelow
\begin{table}[!tbp]
  \centering
  \caption{Default Parameter Values Used.}
  \begin{tabular}{|| p{.12\linewidth}  |  p{.45\linewidth}  | p{.23\linewidth} ||}
    \hline
    Variable & Definition & Value \\ \hline \hline
    $\gls{r0minHex}$ & Cell centers distance  & $[5 \cdot 10^{-2}, 5]\SI{}{\,[\meter]}$ \\ \hline
    $\gls{SRX}$ & Receiver radius  & $ \frac{c}{2}\SI{}{\,[\meter]}$ \\[-0.1cm]
    & \scriptsize{(neighboring receivers touch each other)}&   \\ \hline
    $\gls{LRX}$ & Receiver length   & $\SI{0.2}{\,[\meter]}$ \\ \hline
    $\gls{VRX}$ & Receiver volume   & $\gls{SRX}^2 \pi \gls{LRX} \SI{}{\,[\meter^3]}$ \\ \hline
    $d$ &  Distance between planes  & $\SI{0.5}{\,[\meter]}$\\ \hline
    $\gls{flow}$ &  Flow velocity   & $\SI{0.2}{\,[\meter \per \second]}$ \\ \hline
    $\gls{diffusion}$ &  Diffusion coefficient   & $\SI{0.01}{\,[\meter^2 \per \second]}$ \\ \hline
    $\gls{nrMol}$ &  Number of released molecules   & $100$ \\ \hline
    $\Delta t$ & Time resolution \ac{PBS} \& Monte Carlo Simulation & $10^{-3}\SI{}{\,[\second]}$ \\ \hline
    $T_{\mathrm{sim}}$ & Simulation length & $15 \SI{}{\,[\second]}$ \\ \hline
  \end{tabular}\vspace*{-4mm}
  \label{Table:Parameter}
\end{table}
The default values of the channel parameters are given in \Table{Table:Parameter}.
For the computation of the \ac{CIR}, the infinite sum in \Equation{eq:math_ir_general} was truncated to $21$ terms, i.e., $k_{\mathrm{max}} = 20$. Furthermore, the threshold value \gls{threshold} in \Equation{eq:math_section_threshold_definition} and the \ac{BER} in \Equation{eq:performance:error_prob} were computed for a truncated number of interferers, namely $\gls{nrTrans}-1 = 36$. The chosen numbers of interferers correspond to the first three rings within the hexagonal grid shown in \Figure{graphic:System_model:Grid_hex}.
\scaleSubsubsection
\subsubsection{Monte Carlo Simulation}
\scaleSubsubsectionBelow
To verify the analytical expressions and to show that limiting the number of interferers for numerical evaluation does not affect accuracy, we used Monte Carlo simulation. For Monte Carlo simulation, we considered $\gls{nrTrans}-1 = 1260$ interferers. First, we randomly generated $I = 10^5$ realizations of possible transmit symbol vectors $\boldsymbol{s}_{\textnormal{all}} = [\gls{ss}, \gls{siui}]$. Then, we generated the number of molecules observed at \gls{RX0} based on the channel model in \Equation{eq:system_model_received_molecules}, where RVs  $\gls{cs}$ and $\gls{ciui}$ were modelled as Poisson distributed according to \Equations{equation:math_sec:ss_pois}{equation:math_sec:siui_pois}. Parameters $\gls{csmean}$ and $\gls{cmeaniui}$ were obtained from the \ac{CIR} in \Equation{eq:math_ir_general}. Next, \Equation{eq:performance:error_prob} was numerically evaluated, i.e., the \ac{BER} was computed for all possible threshold values up to $200$, i.e., $\gls{threshold}\leq 200$, and the lowest \ac{BER} together with the corresponding threshold value and error probabilities $p$ and $q$ were selected.
\scaleSubsubsection
\subsubsection{Particle-Based Simulation}
\scaleSubsubsectionBelow
To verify the accuracy of the analytical expressions for the \ac{CIR} in \Equation{eq:math_ir_general}, \ac{3D} stochastic particle-based computer simulations were carried out. The results from \ac{PBS} were averaged over 3000 realizations.
%
%
\scaleSubsection
\subsection{Verification of \ac{CIR}} \label{sec:CIR_verification}
\scaleSubsectionBelow
\begin{figure}[!tbp]
  \begin{minipage}[t]{0.47\textwidth}
    \includegraphics[width=0.95\textwidth]{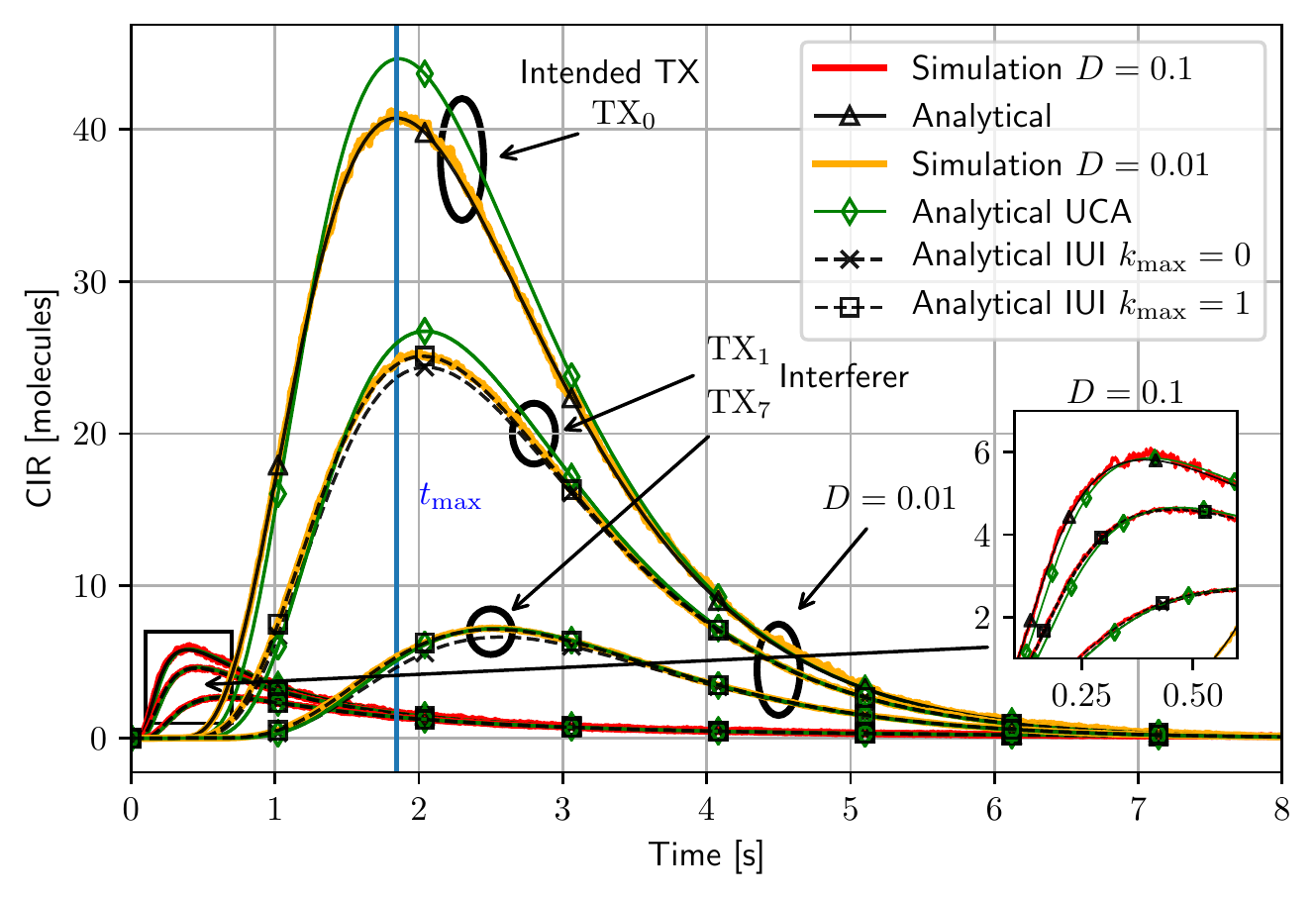}\vspace{-1.5mm}
    \caption{\ac{CIR} for $\ac{TX}_0$, $\ac{TX}_1$, and $\ac{TX}_7$ with $\gls{r0minHex} = 0.2$ for different diffusion coefficients \gls{diffusion}. The proposed analytical results are shown with markers and are compared to results from \ac{PBS}.}
    \label{graphic:evaluation:CIR_validation}
  \end{minipage}
  \vspace{-6mm}
\end{figure}
 \Figure{graphic:evaluation:CIR_validation} shows the \ac{CIR} obtained from the proposed analytical expression \eqref{eq:math_ir_general} (black color) and \ac{PBS} (yellow and red color) as ground truth. Results for the truncation of the sum in \Equation{eq:math_ir_general} to $1$ and $2$ terms, i.e., $k_{\mathrm{max}}=0$ and $k_{\mathrm{max}}=1$, are depicted with dashed lines. Furthermore, we show results for a \ac{CIR} obtained based on the \ac{UCA} at the \ac{RX} (green color), where the molecule concentration at the center of the \ac{RX} is scaled with the volume of the \ac{RX} instead of integrating the concentration over the \ac{RX} volume, i.e., $\mathrm{CIR}_{\mathrm{UCA},i} =  C(x_{\mathrm{RX}}, x_{0,\gls{curTrans}}, y_{\mathrm{RX}}, y_{0,\gls{curTrans}}, z_{\mathrm{RX}}, z_{0,\gls{curTrans}}, t, t_{0,\gls{curTrans}})  \gls{VRX} $  \cite{Jamali2019ChannelMF}. The sampling time $\gls{samplingTime}$ is denoted by $t_{\mathrm{max}}$ (blue color). The verification of the \ac{CIR} is exemplarily done for $\ac{TX}_0$, $\ac{TX}_1$, and $\ac{TX}_7$ for a grid with cell-center distance $\gls{r0minHex} = 0.2$, see \Figure{graphic:System_model:Grid_hex}, and two different diffusion coefficients $D$.

We first concentrate on the \ac{CIR}s for diffusion coefficient $D = 0.01$. From \Figure{graphic:evaluation:CIR_validation}, we observe that for the desired transmitter \gls{TX0}, the proposed \ac{CIR} is in excellent agreement with the PBS results (yellow color). Note that the sum in \eqref{eq:math_ir_general} simplifies for \gls{TX0} to the term $\gamma(1,\frac{\gls{SRX2}}{4 \gls{diffusion} t})$ as $r_{0,0}^2 = 0$ for \gls{TX0}. For $\ac{TX}_1$ and $\ac{TX}_7$ the proposed model in \Equation{eq:math_ir_general} matches the \ac{PBS} result not yet for $k_{\mathrm{max}}=0$, but for $k_{\mathrm{max}}=1$, i.e., for truncating the infinite sum to two terms. The \ac{CIR} based on the \ac{UCA} deviates from the \ac{PBS} result, especially for $\ac{TX}_0$ and $\ac{TX}_1$. The \ac{CIR} expression for the \ac{UCA} is in general simpler, but less precise. The aforementioned observations are also valid for the \ac{CIR}s for $D = 0.1$. Therefore, we conclude that the derived expression for the \ac{CIR} in \eqref{eq:math_ir_general} is accurate. Note that it is sufficient to truncate the infinite sum in \eqref{eq:math_ir_general} after two terms for the parameters considered in \Figure{graphic:evaluation:CIR_validation}. However, we use $k_{\mathrm{max}}=20$ for our analysis in \Section{eval:BER_ARE_Results}.
%
%
\scaleSubsection
\subsection{Evaluation of BER and ARE}\label{eval:BER_ARE_Results}
\scaleSubsectionBelow
\begin{figure}[!tbp]
    \begin{minipage}[t]{0.47\textwidth}
        \includegraphics[width=0.9\textwidth]{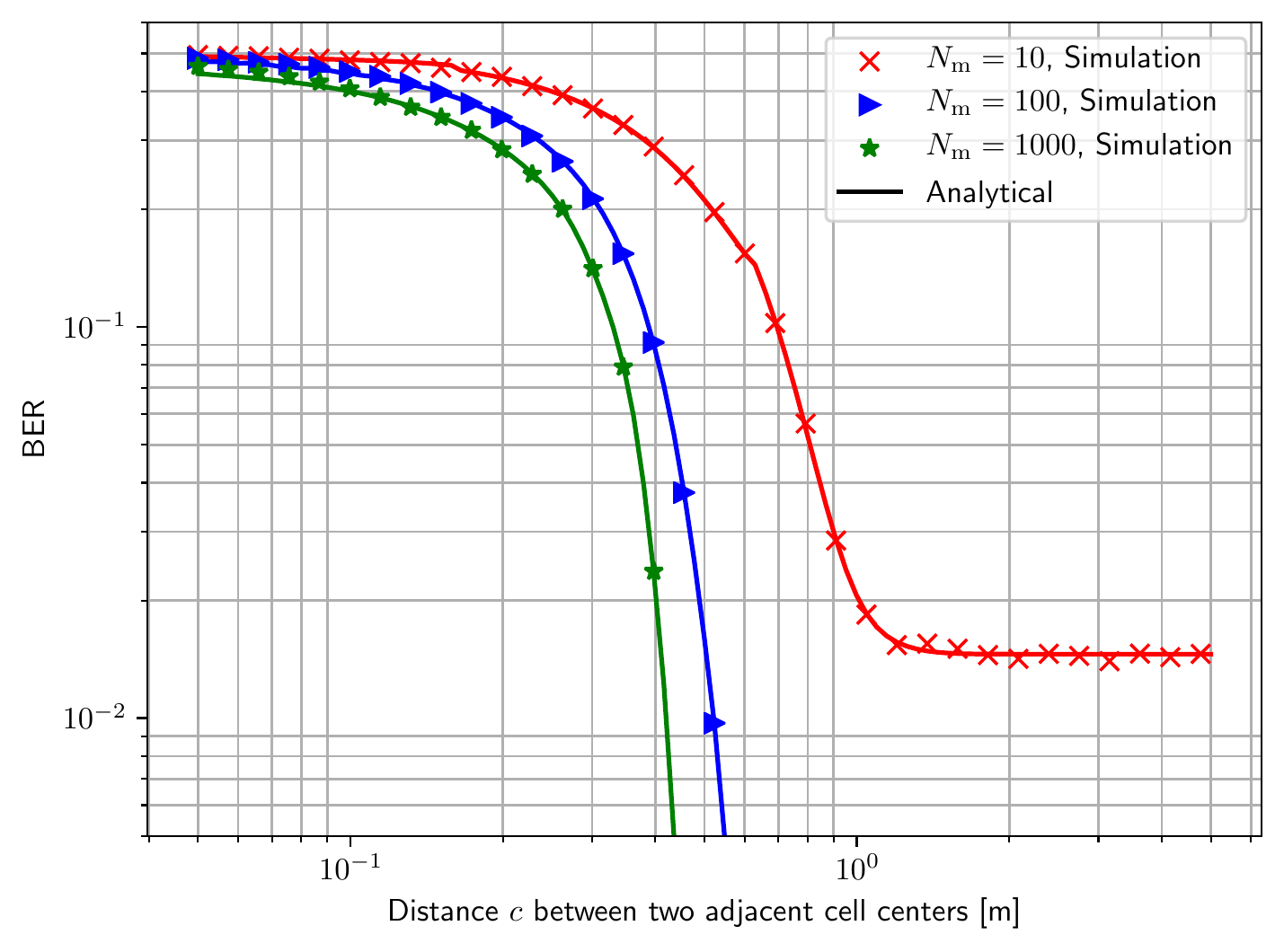}\vspace{-1.5mm}
        \caption{\ac{BER} as a function of cell center distance $\gls{r0minHex}$ for different numbers of released molecules \gls{nrMol}. The results from Monte Carlo simulation are depicted by markers.}
        \label{graphic:evaluation:BER_MolNumber}
    \end{minipage}
    \vspace*{-5mm}
\end{figure}
\begin{figure}[!tbp]
    \begin{minipage}[t]{0.47\textwidth}
        \includegraphics[width=0.9\textwidth]{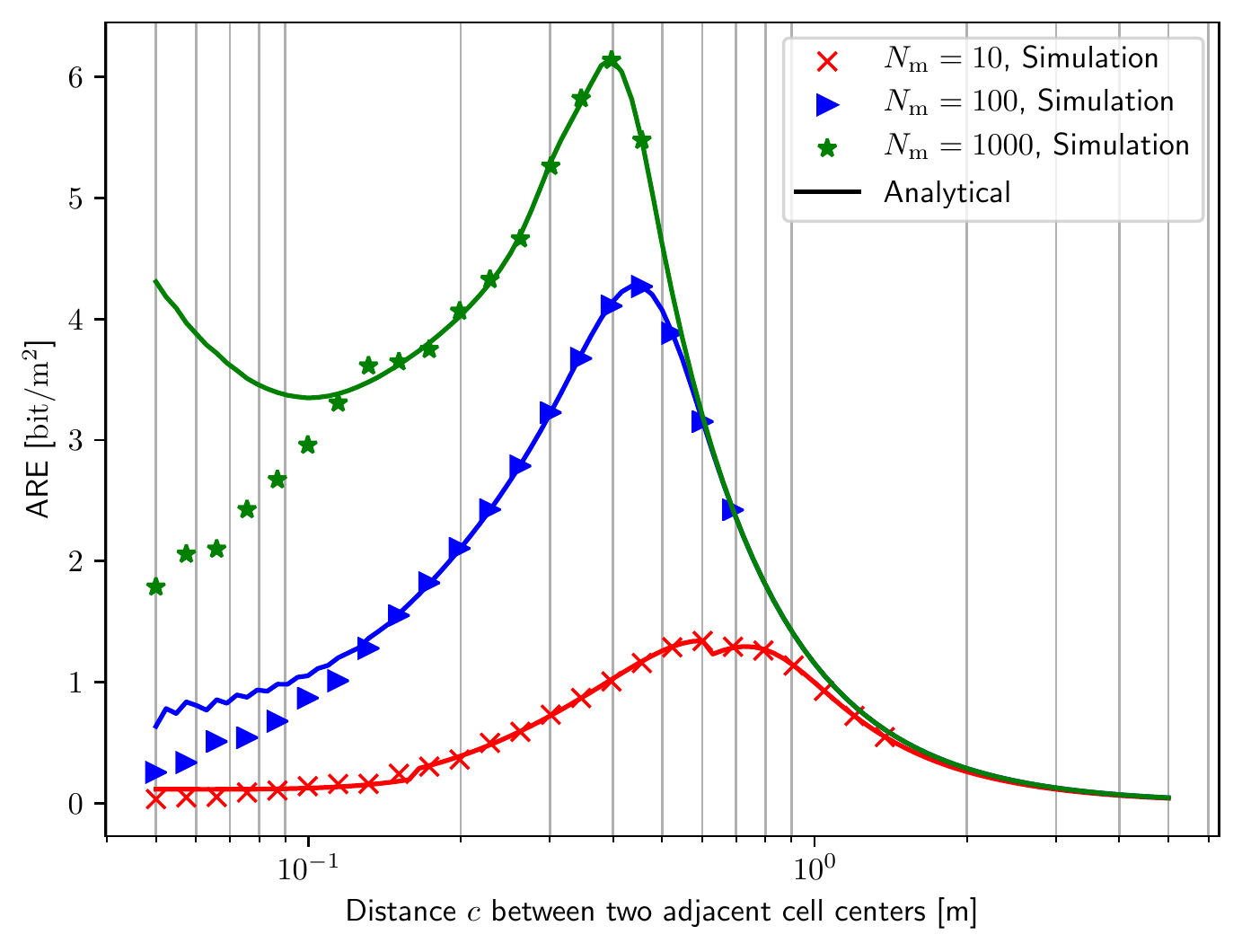}\vspace{-1.5mm}
        \caption{\ac{ARE} as a function of cell center distance $\gls{r0minHex}$ for different numbers of released molecules \gls{nrMol}. The results from Monte Carlo simulation are depicted by markers.}
        \label{graphic:evaluation:rateMolNumber}
    \end{minipage}
     \vspace*{-5mm}
\end{figure}
In \Figures{graphic:evaluation:BER_MolNumber}{graphic:evaluation:rateMolNumber}, \ac{BER} and \ac{ARE} are shown as functions of the hexagonal cell center distance $\gls{r0minHex}$ for different numbers of released molecules \gls{nrMol}.
\Figure{graphic:evaluation:BER_MolNumber} shows that for small $\gls{r0minHex}$, when the \ac{TXs} are closely spaced, the \ac{BER} approaches $0.5$ because of the excessive interference. We observe that increasing $\gls{r0minHex}$ decreases the \ac{BER} and, for large values of $\gls{r0minHex}$, the \ac{IUI} approaches zero and the \ac{BER} reaches a constant error floor. For the range of \ac{BER} values shown in \Figure{graphic:evaluation:BER_MolNumber}, the error floor is visible only for $\gls{nrMol} = 10$, but also occurs for larger $\gls{nrMol}$ at lower \ac{BER} values.
In particular, in the absence of \ac{IUI}, the optimal detection threshold equals one, i.e., $\gls{threshold} = 1$, and detection errors can only occur for $\gls{ss} = 1$. The corresponding error probability $q$ in \Equation{eq:performance:error_prob} is characterized by the probability of receiving zero molecules at \gls{RX0} although \gls{nrMol} molecules were released by \gls{TX0}. In particular, for large values of $\gls{r0minHex}$, \gls{RX0} spans a large area of the \ac{RX}-plane as \gls{SRX} scales with \gls{r0minHex}. Therefore, the error probability is constant and approximately inversely proportional to the probability of at least one molecule reaching the \ac{RX}.
Furthermore, \Figure{graphic:evaluation:BER_MolNumber} shows that the \ac{BER} decreases for increasing $\gls{nrMol}$. We observe that the difference in \ac{BER} for different $\gls{nrMol}$ decreases for smaller cell distances \gls{r0minHex}.

\Figure{graphic:evaluation:rateMolNumber} shows that the \ac{ARE} has a unique maximum and we denote the corresponding cell center distance as $\gls{r0minHex}^{\mathrm{opt}}$. We observe that both, decreasing and increasing \gls{r0minHex} compared to $\gls{r0minHex}^{\mathrm{opt}}$, decreases the \ac{ARE} asymptotically towards $\ac{ARE} \rightarrow 0$. For small \gls{r0minHex}, the considered system is dominated by excessive \ac{IUI}, i.e., the small $R_{\textnormal{SISO}}$ dominates the \ac{ARE} \Equation{eq:performance:effRate}. However, for large \gls{r0minHex}, the system's usage of the spatial resource is not optimal, as the hexagonal cell area reserved for one \ac{TX}-\ac{RX} link is large, i.e., the small $\gls{spatialRate}$ dominates the \ac{ARE} \Equation{eq:performance:effRate}. From the existence of a maximum \ac{ARE}, we conclude that there exists an optimal \ac{TX}-\ac{RX} link density, which is achieved by \ac{TX} positions with cell center distance $\gls{r0minHex} = \gls{r0minHex}^{\mathrm{opt}}$.
%
Next, we focus on the impact of the numbers of released molecules \gls{nrMol}. \Figure{graphic:evaluation:rateMolNumber} shows that increasing $\gls{nrMol}$ also increases the peak value of the \ac{ARE}. We observe that $\gls{r0minHex}^{\mathrm{opt}}$ is smaller for larger numbers of released molecules \gls{nrMol}, i.e., the optimal density of the independent users is larger for larger \gls{nrMol}. We note that, in practical applications, there might be an upper limit on \gls{nrMol} due to limited resources. We further observe small ripples in the \ac{ARE} for $\gls{nrMol} = 10$ and $\gls{nrMol} = 100$ which are caused by the fact that threshold \gls{threshold} is limited to integer values.
Finally, we observe, that the analytical solution deviates from the Monte Carlo simulation result for small \gls{r0minHex}. The reason for this is that for small \gls{r0minHex}, more than $36$ interferers have to be considered in order to properly model the \ac{IUI}. In fact, the actual number of interferers needed to accurately approximate the system behavior increases with decreasing \gls{r0minHex} and should be chosen carefully.

\scaleSection
\section{Conclusion}
\label{sec:conclusion}
In this paper, we focused on the spatial dimension of \ac{MC} systems. We considered a \ac{3D} system with multiple independent and spatially distributed point-to-point transmission links, where the \ac{TXs} and the \ac{RXs} were positioned according to a regular hexagonal pattern, respectively. We developed an analytical expression for the \ac{CIRs} of all existing \ac{TX}-\ac{RX} links. We further applied threshold-based detection and analyzed the \ac{BER} of a single transmission link. The corresponding detection threshold was derived taking into account the statistics of the MC channel and the statistics of the IUI. Next, we proposed the \ac{ARE} as a new performance metric for \ac{MC} systems to characterize how efficiently given \ac{TX} and \ac{RX} areas are used for information transmission. The \ac{ARE} reflects the impact of the user density, i.e., the number of \ac{TXs} and \ac{RXs} per unit area, and the \ac{IUI} dependent performance of each individual transmission link on the overall performance of the \ac{MC} system. Finally, quantitative results for the optimal user density for maximization of the \ac{ARE} of \ac{MC} systems were provided.

The \ac{ARE} analyzed in this paper provides a new perspective for designing multiuser \ac{MC} systems. Although we have considered a particular multiuser \ac{MC} setup, the concept to analyze the system performance in terms of the \ac{ARE} can be extended to other \ac{MC} systems, including different \ac{MC} channels, different \ac{TX} and \ac{RX} grids, and different types of \ac{TXs} and \ac{RXs}.
%

\vspace*{-0.2cm}
\bibliographystyle{ACM-Reference-Format}
\bibliography{literature}
\end{document}